\documentclass[preprint,showpacs,preprintnumbers,amsmath,amssymb,superscriptaddress]{revtex4}
\usepackage{graphicx}
\usepackage{dcolumn}
\usepackage{amssymb}
\usepackage{amsmath}
\usepackage{bm}
\usepackage[ulem=normalem]{changes}
\usepackage{hyperref}
\usepackage{xfrac}

\begin{document}
\title{Relation between E2 transitions in even--even and odd-mass nuclei}

\author{P.~Van~Isacker}
\affiliation{Grand Acc\'el\'erateur National d'Ions Lourds,
CEA/DRF - CNRS/IN2P3, Bvd Henri Becquerel, F-14076 Caen, France}

\date{\today}

\begin{abstract}
How is the \mbox{$B({\rm E2};2_1^+\rightarrow0_1^+)$} value in an even--even nucleus
related to corresponding \mbox{$B({\rm E2};J_{\rm i}\rightarrow J_{\rm f})$} values
in a neighbouring odd-mass nucleus?
If either neutrons or protons are confined to a single-$j$ orbital
and if the nucleon--nucleon interaction conserves seniority,
a simple relation between the two properties is obtained,
which may differ from what is found
in the weak-coupling limit of the core--particle model.
This single-$j$ relation is substantially perturbed
if several non-degenerate orbitals are considered.
An application to recently measured $B$(E2) values in neutron-deficient tin isotopes is presented.
\end{abstract}

\maketitle

\section{Introduction}
\label{s_intro}
All even--even nuclei have a ground state with angular momentum and parity \mbox{$J^\pi=0^+$}
and most have a first-excited level with \mbox{$J^\pi=2^+$}.
Much can be learned about the properties of nuclei
from the study of these two nuclear states.
In particular, a low excitation energy of the $2_1^+$ level
and a concomitant large electric quadrupole (E2) \mbox{$2_1^+\rightarrow0_1^+$} transition 
are indications of (quadrupole) deformation.
Semi-magic nuclei, with either neutrons or protons in the valence shell,
typically exhibit a rather constant $2_1^+$ excitation energy
and a \mbox{$B({\rm E2};2_1^+\rightarrow0_1^+)$} value
that varies in a characteristic parabolic way with particle number---properties
indicative of pairing correlations.

Given the importance of the \mbox{$B({\rm E2};2_1^+\rightarrow0_1^+)$} value
as a structural indicator of even--even nuclei,
it is of interest to study the corresponding E2 property in odd-mass nuclei.
This is the purpose of this contribution.
I start with a more precise outline of the problem.

\section{Outline of the problem}
\label{s_outline}
Let us assume,
as in the particle-core coupling model of de-Shalit~\cite{Shalit61},
that the low-energy levels in the odd-mass nucleus
can be described as a core state (either $0_1^+$ or $2_1^+$)
coupled to a nucleon in a $j$ orbital.
Under this assumption the ground state of the odd-mass nucleus is written as \mbox{$|0_1^+\times j;J=j\rangle$}
while the multiplet of states $|j-2|,\dots,j+2$,
arising from the coupling of the nucleon to the $2_1^+$ level of the core, is \mbox{$|2_1^+\times j;J\rangle$}.
In this approximation the E2 matrix element in the odd-mass nucleus is given by
\begin{equation}
\langle0_1^+\times j;J_{\rm f}\!=\!j\|\hat T({\rm E2})\|2_1^+\times j;J_{\rm i}\rangle=
\sqrt{\frac{2J_{\rm i}+1}{5}}\langle0_1^+\|\hat T({\rm E2})\|2_1^+\rangle.
\label{e_relme2}
\end{equation}
%with $[x]\equiv\sqrt{2x+1}$.
This implies a relation between $B$(E2) values in the even--even and odd-mass nuclei,
\begin{equation}
B({\rm E2};J_{\rm i}\rightarrow J_{\rm f}=j)=B({\rm E2};2_1^+\rightarrow0_1^+),
\label{e_relbe2}
\end{equation}
which in turn implies $\sum B({\rm E2}\uparrow)=B({\rm E2};0_1^+\rightarrow2_1^+)$,
where the sum is from the ground state of the odd-mass nucleus
to all members of the \mbox{$2_1^+\times j$} multiplet.

The E2 sum rule,
implicit in the particle-core description of the odd-mass nucleus,
has been tested in a number of semi-magic even--even nuclei
and their odd-mass neighbours.
The first case was reported by Tuttle {\it et al.}~\cite{Tuttle76}
by comparing E2 strength in $^{113,115}$In with that in $^{114,116}$Sn.
Subsequent examples generally confirmed the validity of the E2 sum rule
with some notable exceptions, see Gray {\it et al.}~\cite{Gray20} and references therein.
In all nuclei reported in Ref.~\cite{Gray20} the odd particle's type ({\it i.e.}, its isospin projection)
is different from that of the valence nucleons in the semi-magic even-even nucleus.
Only recently, by way of lifetime measurements in $^{105}$Sn,
a case was established where all valence nucleons are of the same type~\cite{Pasqualato23}.

It is expected that the relation~(\ref{e_relbe2})
is only approximately satisfied in the shell model.
The reasons for this are two-fold.
If the odd particle's type is different from that of the other valence nucleons,
the neutron-proton interaction may lead to an increased collectivity in the odd-mass nucleus
and a concomitant violation of the E2 sum rule~\cite{Gray20}.
If, on the other hand, all valence nucleons are of the same type,
deviations from the relation~(\ref{e_relbe2}) arise due to the Pauli principle,
which imposes anti-symmetry between the odd particle and the other nucleons.
%and are less dependent on the nucleon--nucleon interaction.
In the simple case of identical nucleons in a single-$j$ orbital,
the wave function of the odd-mass state $|J_{\rm i}\rangle$
involves a sum over {\em all} core states $2,4,\dots$
weighted by coefficients of fractional parentage (CFPs).
As a result the odd-mass \mbox{$B({\rm E2};J_{\rm i}\rightarrow J_{\rm f})$} value
can be larger or smaller compared to the even--even \mbox{$B({\rm E2};2_1^+\rightarrow0_1^+)$} value,
depending on the value of $J_{\rm i}$.

In this paper the relation
between the \mbox{$B({\rm E2};J_{\rm i}\rightarrow J_{\rm f})$}
and \mbox{$B({\rm E2};2_1^+\rightarrow0_1^+)$} values,
hence\-forth referred to as the odd--even E2 relation, is discussed.
Throughout it is assumed that there are either neutrons or protons in the valence shell,
corresponding to the second case mentioned above.
A simple odd--even E2 relation for a single-$j$ orbital
is derived in Sect.~\ref{s_singlej} under the assumption of conservation of seniority.
As shown in Sect.~\ref{s_tin} with the example of the tin isotopes,
the presence of several non-degenerate orbitals
has a major influence on the odd--even E2 relation
and will be crucial when comparing to data.
Conclusions are presented in Sect.~\ref{s_con}.

\section{The odd--even E2 relation in a single-$j$ orbital}
\label{s_singlej}
For identical nucleons in a single-$j$ orbital,
it is known that seniority is conserved to a good approximation~\cite{Shalit63,Talmi93}.
The $0_1^+$ ground state of an even--even nucleus has (approximately) seniority $\upsilon=0$
while the $2_1^+$ state carries \mbox{$\upsilon=2$}.
It is then possible to relate the \mbox{$B({\rm E2};2_1^+\rightarrow0_1^+)$} value
in the $n$-nucleon system to that of the two-nucleon system,
\begin{equation}
B({\rm E2};j^n2_1\rightarrow j^n0_1)=
%\nonumber\\&\quad=
\frac{n(2j+1-n)}{2(2j-1)}
B({\rm E2};j^22_1\rightarrow j^20_1),
\;n\;{\rm even}.
\label{e_be2e}
\end{equation}
In odd-mass nuclei one considers
E2 transitions from levels with angular momentum $J_{\rm i}$ and seniority \mbox{$\upsilon=3$}
to the ground state with \mbox{$J_{\rm f}=j$} and \mbox{$\upsilon=1$},
for which the corresponding formula reads
\begin{equation}
B({\rm E2};j^nJ_{\rm i}\rightarrow j^nJ_{\rm f})=
%\nonumber\\&\quad=
\frac{(n-1)(2j-n)}{2(2j-3)}
B({\rm E2};j^3J_{\rm i}\rightarrow j^3J_{\rm f}),
\;n\;{\rm odd}.
\label{e_be2o}
\end{equation}
It is assumed henceforth in this section that \mbox{$J_{\rm f}=j$}.

If the $j$ orbital is less than half-filled, \mbox{$n\leq(2j+1)/2$},
one relates the \mbox{$B({\rm E2};J_{\rm i}\rightarrow J_{\rm f})$} value
in the odd-mass nucleus with \mbox{$n+1$} nucleons
to the \mbox{$B({\rm E2};2_1^+\rightarrow0_1^+)$} value
in the even--even nucleus with $n$ nucleons.
The combination of Eqs.~(\ref{e_be2e}) and~(\ref{e_be2o})
leads to the relation
\begin{equation}
\frac{B({\rm E2};j^{n+1}J_{\rm i}\rightarrow j^{n+1}J_{\rm f})}
{B({\rm E2};j^n2_1\rightarrow j^n0_1)}=
%\nonumber\\&\quad=
\frac{(2j-1)(2j-1-n)}{(2j-3)(2j+1-n)}
\frac{B({\rm E2};j^3J_{\rm i}\rightarrow j^3J_{\rm f})}
{B({\rm E2};j^22_1\rightarrow j^20_1)}.
\label{e_be2eo1}
\end{equation}
If the $j$ orbital is more than half-filled, \mbox{$n\geq(2j+1)/2$},
one relates the \mbox{$B({\rm E2};J_{\rm i}\rightarrow J_{\rm f})$} value
in the odd-mass nucleus with \mbox{$n-1$} nucleons
to the \mbox{$B({\rm E2};2_1^+\rightarrow0_1^+)$} value
in the even--even nucleus with $n$ nucleons.
In this case one finds from Eqs.~(\ref{e_be2e}) and~(\ref{e_be2o})
\begin{equation}
\frac{B({\rm E2};j^{n-1}J_{\rm i}\rightarrow j^{n-1}J_{\rm f})}
{B({\rm E2};j^n2_1\rightarrow j^n0_1)}=
%\nonumber\\&\qquad=
\frac{(2j-1)(n-2)}{(2j-3)n}
\frac{B({\rm E2};j^3J_{\rm i}\rightarrow j^3J_{\rm f})}
{B({\rm E2};j^22_1\rightarrow j^20_1)}.
\label{e_be2eo2}
\end{equation}
Note that relations~(\ref{e_be2eo1}) and~(\ref{e_be2eo2})
can be obtained from each other through the substitution \mbox{$n\rightarrow2j+1-n$},
as should be since the problem is invariant under particle--hole conjugation~\cite{Lawson80}.

In view of Eqs.~(\ref{e_be2eo1}) and~(\ref{e_be2eo2})
the problem of finding the odd--even E2 relation in a single-$j$ orbital
is reduced to that of finding the relation between the two- and three-nucleon systems.
This problem was discussed by Karayonchev {\it et al.}~\cite{Karayonchev19},
whose method is applied here for a one-body E2 operator.

First I recall a few properties of anti-symmetric three-nucleon states~\cite{Shalit63,Talmi93}.
A three-nucleon state can be written as \mbox{$|j^2(I)j;J\rangle$},
where two nucleons are first coupled to angular momentum $I$,
which is subsequently coupled with the third nucleon to total angular momentum $J$.
This state is not anti-symmetric in all three nucleons;
it can be made so by applying the anti-symmetry operator $\cal A$,
\begin{equation}
{\cal A}|j^2(I)j;J\rangle
\propto|j^3[I]J\rangle=
\sum_R
%c^{j^2R}_{j^3[I]J}\;
[j^2(R)jJ|\}j^3[I]J]\;
|j^2(R)j;J\rangle,
\label{e_state}
\end{equation}
where \mbox{$[j^2(R)jJ|\}j^3[I]J]$} is a \mbox{$3\rightarrow2$} CFP.
The square bracket $[I]$ labels the three-nucleon state
and indicates that it has been obtained after anti-symmetrisation of \mbox{$|j^2(I)j;J\rangle$}.
The states \mbox{$|j^3[I]J\rangle$} are normalised and define a non-orthogonal and overcomplete basis
so that in general the states with \mbox{$I=0,2,\dots,2j-1$} are not independent.

A related basis \mbox{$|j^3\upsilon\alpha_IJ\rangle$}
is defined through the seniority quantum number $\upsilon$,
which counts the number of nucleons not in pairs coupled to zero.
The state with seniority \mbox{$\upsilon=1$} always exists and is unique,
\mbox{$|j^3\upsilon=1,J\rangle$}${}={}$\mbox{$|j^3[I=0]J\rangle$}.
All other three-nucleon states have seniority $\upsilon=3$
and originate from the anti-symmetrisation of some state \mbox{$|j^2(I)j;J\rangle$} with \mbox{$I\neq0$}.
For \mbox{$J\neq j$} this relation is direct,
\mbox{$|j^3\upsilon=3,\alpha_IJ\rangle$}${}={}$\mbox{$|j^3[I\neq0]J\rangle$}.
A label $\alpha_I$, additional to seniority $\upsilon$, is needed
if several linearly independent states with seniority \mbox{$\upsilon=3$} exist,
as may happen for \mbox{$j\geq\sfrac{11}{2}$}.
In that case linearly independent states can be constructed by considering different values of $I$.
While the first state can be defined by direct association with the basis \mbox{$|j^3[I]J\rangle$},
other states must be constructed with a Gram--Schmidt orthogonalisation procedure.
In the subsequent discussion seniority \mbox{$\upsilon=3$} states
are constructed from \mbox{$|j^2(I=2)j;J\rangle$} unless otherwise stated.
This is motivated by the original objective of relating the seniority basis
to the nucleon--core coupled states
\mbox{$|0_1^+\times j;J=j\rangle$} and \mbox{$|2_1^+\times j;J\rangle$}.
Since other seniority \mbox{$\upsilon=3$} states will not be considered,
there is no need for a Gram--Schmidt orthogonalisation if \mbox{$J\neq j$}.
For $J=j$ the Gram--Schmidt orthogonalisation of seniority \mbox{$\upsilon=3$} states cannot be avoided
since the states \mbox{$|j^3[I=0]J=j\rangle$} and \mbox{$|j^3[I\neq0]J=j\rangle$} are not orthogonal.

The above results can be summarised by giving the correspondence
between CFPs in the two bases \mbox{$|j^3\upsilon\alpha_IJ\rangle$} and \mbox{$|j^3[I]J\rangle$},
%\begin{align}
%c^{j^2R}_{j^3\upsilon=1,J}&{}=c^{j^2R}_{j^3[0]J},
%\;J=j,
%\label{e_cfpcor}\\
%c^{j^2R}_{j^3\upsilon=3,\alpha_IJ}&{}=c^{j^2R}_{j^3[I]J},
%\;J\neq j\;\&\;I\neq0,
%\nonumber\\
%c^{j^2R}_{j^3\upsilon=3,\alpha_IJ}&{}=
%\frac{c^{j^2R}_{j^3[I]J}-\langle j^3[0]J|j^3[I]J\rangle\;c^{j^2R}_{j^3[0]J}}
%{\sqrt{1-\langle j^3[0]J|j^3[I]J\rangle^2}},
%\;J=j\;\&\;I\neq0,
%\nonumber
%\end{align}
\begin{align}
[j^2(R)jJ|\}j^3\upsilon=1,J]&{}=[j^2(R)jJ|\}j^3[0]J],
\;J=j,
\label{e_cfpcor}\\
[j^2(R)jJ|\}j^3\upsilon=3,\alpha_IJ]&{}=[j^2(R)jJ|\}j^3[I]J],
\;J\neq j\;\&\;I\neq0,
\nonumber\\
[j^2(R)jJ|\}j^3\upsilon=3,\alpha_IJ]&{}=
\frac{[j^2(R)jJ|\}j^3[I]J]-O_{0I}\;[j^2(R)jJ|\}j^3[0]J]}
{\sqrt{1-O_{0I}^2}},
\;J=j\;\&\;I\neq0,
\nonumber
\end{align}
where \mbox{$O_{II'}\equiv\langle j^3[I]J|j^3[I']J\rangle$} is an overlap matrix element.
The relations~(\ref{e_cfpcor}) are valid
as long as only one seniority \mbox{$\upsilon=3$} state is considered for a given $J$
and are useful since the CFPs in the \mbox{$|j^3[I]J\rangle$} basis
are known in closed form~\cite{Shalit63,Talmi93}.

Since the electric quadrupole operator \mbox{$\hat T({\rm E2})$} is taken to be of one-body character,
its reduced matrix elements in the two- and three-nucleon systems are related through
\begin{align}
&\langle j^3\upsilon\alpha J\|\hat T({\rm E2})\|j^3\upsilon'\alpha'J'\rangle=
\frac{3}{2}(-)^{j+J}\sqrt{(2J+1)(2J'+1)}
\nonumber\\&\times
\sum_{RR'\;{\rm even}}
[j^2(R)jJ|\}j^3\upsilon\alpha J]
[j^2(R)jJ|\}j^3\upsilon'\alpha'J]
\biggl\{\!\!\begin{array}{ccc}J&J'&2\\R'&R&j\end{array}\!\!\biggr\}
\langle j^2R\|\hat T({\rm E2})\|j^2R'\rangle.
\label{e_me3a}
\end{align}
Furthermore, the E2 matrix element in the two-nucleon system is known in closed form,
\begin{equation}
\langle j^2R\|\hat T({\rm E2})\|j^2R'\rangle=
%\nonumber\\&\quad=
e\,\left(N+\textstyle{{\frac32}}\right)x_j\sqrt{20(2R+1)(2R'+1)}
\biggl\{\!\!\begin{array}{ccc}j&j&2\\R&R'&j\end{array}\!\!\biggr\}b^2,
\label{e_me2}
\end{equation}
where $e$ is the effective charge of the nucleon,
$N$ is the major oscillator quantum number
and $b$ is the length parameter of the harmonic oscillator,
and with
\begin{equation}
x_j=(-)^{j+1/2}\sqrt{\frac{(2j-1)(2j+1)(2j+3)}{64\pi j(j+1)}}.
\label{e_xfac}
\end{equation}

The application of Eq.~(\ref{e_me3a}) to initial and final states
\mbox{$|j^3\upsilon=3,\alpha_IJ_{\rm i}\rangle$} and \mbox{$|j^3\upsilon=1,J_{\rm f}\rangle$},
in combination with Eq.~(\ref{e_me2})
leads to the following relation between the $B$(E2) values:
\begin{equation}
\frac{B({\rm E2};j^3\upsilon=3,\alpha_IJ_{\rm i}\rightarrow j^3\upsilon=1,J_{\rm f})}
{B({\rm E2};j^22_1\rightarrow j^20_1)}=
q_j(\alpha_IJ_{\rm i}),
\label{e_be23b}
\end{equation}
where \mbox{$q_j(\alpha_IJ_{\rm i})$} is given by
\begin{align}
q_j(\alpha_IJ_{\rm i})&{}=
\frac{45}{4}
\biggl((2j+1)\sum_{RR'\;{\rm even}}\sqrt{(2R+1)(2R'+1)}
\nonumber\\&\quad\times
[j^2(R)jJ|\}j^3\upsilon=3,\alpha_IJ_{\rm i}]
[j^2(R)jJ|\}j^3\upsilon=1,J_{\rm f}]
\biggl\{\!\!\begin{array}{ccc}j&J_{\rm i}&2\\R&R'&j\end{array}\!\!\biggr\}
\biggl\{\!\!\begin{array}{ccc}j&j&2\\R&R'&j\end{array}\!\!\biggr\}\biggr)^2.
\label{e_qcoef1}
\end{align}
The ratio of $B$(E2) values on the right-hand-side of Eqs.~(\ref{e_be2eo1}) and~(\ref{e_be2eo2})
can be replaced by \mbox{$q_j(\alpha_IJ_{\rm i})$},
leading to the following result for the odd--even E2 relations in a single-$j$ orbital:
\begin{align}
\frac{B({\rm E2};j^{n+1}\alpha_IJ_{\rm i}\rightarrow j^{n+1}J_{\rm f})}
{B({\rm E2};j^n2_1\rightarrow j^n0_1)}={}&
\frac{(2j-1)(2j-1-n)}{(2j-3)(2j+1-n)}q_j(\alpha_IJ_{\rm i}),
\nonumber\\
\frac{B({\rm E2};j^{n-1}\alpha_IJ_{\rm i}\rightarrow j^{n-1}J_{\rm f})}
{B({\rm E2};j^n2_1\rightarrow j^n0_1)}={}&
\frac{(2j-1)(n-2)}{(2j-3)n}q_j(\alpha_IJ_{\rm i}),
\label{e_be2eo3}
\end{align}
if the $j$ orbital is less or more than half-filled, respectively.

To recapitulate, the odd--even E2 relations~(\ref{e_be2eo3})
are valid in a single-$j$ orbital for a one-body E2 operator
under the assumption that seniority is a conserved quantum number.
In the even--even nucleus the initial and final states in the transition
have seniority \mbox{$\upsilon=2$} and \mbox{$\upsilon=0$}, respectively;
in the odd-mass nucleus they carry \mbox{$\upsilon=3$} and \mbox{$\upsilon=1$}.

The sum in the coefficient \mbox{$q_j(\alpha_IJ_{\rm i})$} can be carried out analytically.
A lengthy calculation yields the following expression:
\begin{equation}
q_j(\alpha_IJ_{\rm i})=
\frac{(2I+1)(2j+1)}{5(2j-1)}
\frac{r_j(2,I,J_{\rm i})^2}{r_j(I,I,J_{\rm i})},
\label{e_qcoef2}
\end{equation}
which for \mbox{$I=2$} simplifies to
\begin{equation}
q_j(\alpha_2J_{\rm i})=
\frac{2j+1}{2j-1}r_j(2,2,J_{\rm i}),
\label{e_qcoef3}
\end{equation}
with
\begin{equation}
r_j(I,I',J_{\rm i})=
%\nonumber\\&\quad=
\delta_{II'}+10\,\biggl\{\!\!\begin{array}{ccc}j&j&I\\J_{\rm i}&j&I'\end{array}\!\!\biggr\}
-{\displaystyle\frac{20}{(2j-1)(2j+1)}}\delta_{jJ_{\rm i}}.
\label{e_rcoef2}
\end{equation}
The explicit expressions for the coefficients \mbox{$q_j(\alpha_2J_{\rm i})$} are
\begin{align}
q_j(\alpha_2J_{\rm i}=j-2)={}&
\frac{(2j-5)(2j^2+5j+12)}{j(2j-1)^2},
\nonumber\\
q_j(\alpha_2J_{\rm i}=j-1)={}&
\frac{(j+2)(2j-3)(2j^2+j+35)}{j(j+1)(2j-1)^2},
\nonumber\\
q_j(\alpha_2J_{\rm i}=j)={}&
\frac{(j+2)(j+3)(2j-7)(2j-5)(2j-3)}{j(j+1)(2j-1)^2(2j+3)},
\nonumber\\
q_j(\alpha_2J_{\rm i}=j+1)={}&
\frac{(j+3)(j+4)(2j-5)(2j-3)}{j(j+1)(2j-1)(2j+3)},
\nonumber\\
q_j(\alpha_2J_{\rm i}=j+2)={}&
\frac{(2j-3)(2j^2+9j+19)}{(j+1)(2j-1)(2j+3)}.
\label{e_qcoef4}
\end{align}

\begin{table}
\centering
\caption{The coefficients \mbox{$q_j(\alpha_2J_{\rm i})$} in Eq.~(\ref{e_be2eo3}).}
\label{t_qcoef}
\begin{tabular}{l|cccccccccccc}
\hline\hline
&&$j=\sfrac{5}{2}$
&~~~&$j=\sfrac{7}{2}$
&~~~&$j=\sfrac{9}{2}$
&~~~&$j=\sfrac{11}{2}$
&~~~~~&$j=\sfrac{13}{2}$
&~~~~~&$j=\sfrac{15}{2}$\\
%$J_{\rm i}$&&&&&&\\
\hline\\[-5mm]
$J_{\rm i}=j-2$
&&---&&
$\frac{6}{7}$&&
$\frac{25}{24}$&&
$\frac{12}{11}\phantom{^*}$&&
$\frac{43}{39}^*$&&
$\frac{54}{49}^*$\\[1mm]
$J_{\rm i}=j-1$
&&$\frac{45}{14}$
&&$\frac{22}{9}$
&&$\frac{65}{33}$
&&$\frac{1212}{715}^*$
&&$\frac{119}{78}^*$
&&$\frac{1178}{833}^*$\\[1mm]
$J_{\rm i}=j$
&&---
&&---
&&$\frac{65}{528}$
&&$\frac{1224}{5005}\phantom{^*}$
&&$\frac{323}{936}\phantom{^*}$
&&$\frac{152}{357}^*$\\[1mm]
$J_{\rm i}=j+1$
&&---
&&$\frac{26}{63}$
&&$\frac{85}{132}$
&&$\frac{3876}{5005}\phantom{^*}$
&&$\frac{133}{156}^*$
&&$\frac{46}{51}^*$\\[1mm]
$J_{\rm i}=j+2$
&&$\frac{27}{28}$
&&$\frac{10}{9}$
&&$\frac{25}{22}$
&&$\frac{516}{455}^*$
&&$\frac{9}{8}^*$
&&$\frac{398}{357}^*$\\[1mm]
\hline\hline
\multicolumn{7}{c}{$^*$Several states exist with $\upsilon=3$ and $J_{\rm i}$.}
\end{tabular}
\end{table}
The coefficients \mbox{$q_j(\alpha_2J_{\rm i})$} are tabulated in Table~\ref{t_qcoef}
for the cases of interest in nuclei.
For \mbox{$j\leq\sfrac{7}{2}$} not all values \mbox{$J_{\rm i}=j-2,\dots,j+2$}
are allowed among the seniority \mbox{$\upsilon=3$} states.
This follows from the explicit expressions~(\ref{e_qcoef4}).
For example, for \mbox{$j=\sfrac{3}{2}$} no seniority \mbox{$\upsilon=3$} state exists
and hence all coefficients \mbox{$q_j(\alpha_2J_{\rm i})$} contain a factor \mbox{$2j-3$} in the numerator.
Furthermore, no seniority \mbox{$\upsilon=3$} state with \mbox{$J_{\rm i}=j$}
exists for \mbox{$j=\sfrac{5}{2}$} and \mbox{$j=\sfrac{7}{2}$}
and therefore \mbox{$q_j(\alpha_2J_{\rm i}=j)$}
also contains the factors \mbox{$2j-5$} and \mbox{$2j-7$} in the numerator.
For \mbox{$j\leq\sfrac{9}{2}$} at most one \mbox{$\upsilon=3$} state occurs
for a given angular momentum $J_{\rm i}$.
As a consequence its structure is independent of $I$
and so is the expression for the coefficient \mbox{$q_j(\alpha_IJ_{\rm i})$} in Eq.~(\ref{e_qcoef2}).
This is no longer true if several states exist with seniority \mbox{$\upsilon=3$} and angular momentum $J_{\rm i}$
(indicated with an asterisk in Table~\ref{t_qcoef}),
in which case \mbox{$q_j(\alpha_IJ_{\rm i})$} does depend on $I$.

The explicit expressions~(\ref{e_qcoef4}) for the coefficients \mbox{$q_j(\alpha_2J_{\rm i})$} show that
\begin{equation}
\lim_{j\rightarrow\infty}q_j(\alpha_2J_{\rm i})=1.
\label{e_qcoeflim}
\end{equation}
Therefore, the property~(\ref{e_relbe2}) is satisfied in the large-$j$ limit.
Nevertheless, important deviations from this limit occur for finite values of $j$,
as can be seen from Table~\ref{t_qcoef}.
In particular, the \mbox{$B({\rm E2};J_{\rm i}\rightarrow J_{\rm f})$} value
is larger than the neighbouring \mbox{$B({\rm E2};2_1\rightarrow0_1)$} value for \mbox{$J_{\rm i}=j-1$}
while it is smaller for \mbox{$J_{\rm i}=j$} and \mbox{$J_{\rm i}=j+1$}.

\section{The example of the tin isotopes}
\label{s_tin}
Over the last 20 years considerable experimental effort has been devoted
to the determination of \mbox{$B({\rm E2};2_1^+\rightarrow0_1^+)$} values
in neutron-deficient even--even tin isotopes,
mainly by Coulomb excitation in inverse kinematics~\cite{Vaman07,Ekstrom08,Guastalla13,Bader13,Doornenbal14}
and, more recently, also via lifetime measurements~\cite{Siciliano20}.
These results are of interest since they reveal the amount of quadrupole collectivity in semi-magic $^A$Sn nuclei
as the doubly-magic nucleus $^{100}$Sn is approached.
While there still exists considerable scatter of the experimental results,
it is by now clear that the extracted $B$(E2) values are larger than
what is expected from a shell-model description with only active neutrons in the \mbox{50--82} shell
and that proton excitations across the \mbox{$Z=50$} shell gap
play an essential role in the observed enhancement of quadrupole collectivity.

Very recently the same problem was studied from a different perspective
by measuring lifetimes in the odd-mass isotope $^{105}$Sn~\cite{Pasqualato23}.
Specifically, the authors of that study report on the lifetime of the $(\sfrac{11}{2}^+)$ level at 1394~keV,
which in first approximation can be thought of as a neutron in the $0g_{7/2}$ orbital
coupled to the $2_1^+$ excitation of $^{104}$Sn.
Its lifetime is essentially determined by $\gamma$ decay to the $(\sfrac{7}{2}^+)$ level at 200~keV,
which can be interpreted as the coupling of a neutron in the $0g_{7/2}$ orbital to the ground state of $^{104}$Sn.
As argued in the previous sections,
the \mbox{$B({\rm E2};2_1^+\rightarrow0_1^+)$}
and \mbox{$B({\rm E2};\sfrac{11}{2}_1^+\rightarrow\sfrac{7}{2}_1^+)$} values
are related (although not in a trivial fashion)
and therefore measurements of E2 strength in odd-mass neutron-deficient tin isotopes
can shed light on the question of quadrupole collectivity.

The dominant neutron orbitals in the neutron-deficient tin isotopes are $1d_{5/2}$ and $0g_{7/2}$;
it is unlikely that any of the two orbitals on its own can explain the observations.
This is obvious for $1d_{5/2}$ since identical nucleons in this orbital
cannot couple to the angular momenta \mbox{$J=\sfrac{11}{2}$} or \mbox{$J=\sfrac{7}{2}$}.
States with angular momentum \mbox{$J=\sfrac{11}{2}$} and \mbox{$J=\sfrac{7}{2}$}
exist for an odd number of nucleons in $0g_{7/2}$.
For \mbox{$j=\sfrac{7}{2}$} the ratio (see Table~\ref{t_qcoef}),
\begin{equation}
\frac{B({\rm E2};j^3\sfrac{11}{2}_1\rightarrow j^3\sfrac{7}{2}_1)}
{B({\rm E2};j^22_1\rightarrow j^20_1)}=\frac{10}{9},
\label{e_ratio}
\end{equation}
predicts a slightly larger $B$(E2) value in the odd-mass nuclei,
which is at variance with the observed trend.
The simplest possible approach is therefore to consider both orbitals, $1d_{5/2}$ and $0g_{7/2}$.
This can only be a reasonable approximation if this shell-model space is sufficiently isolated.
Since proton excitations across the \mbox{$Z=50$} shell closure are not included in this analysis,
one cannot expect to obtain close agreement with the data.

\begin{figure}
\centering
\includegraphics[width=8cm]{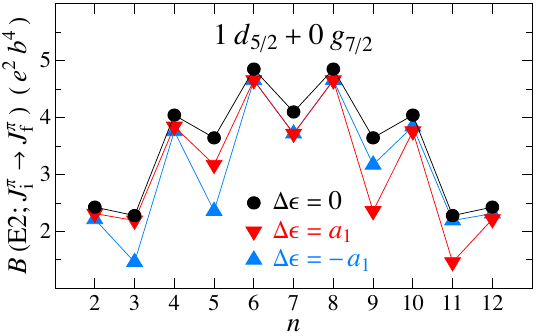}
\caption{Evolution with nucleon number $n$ of E2 transition strength
in the shell-model space consisting of the $1d_{5/2}$ and $0g_{7/2}$ orbitals.
The Hamiltonian~(\ref{e_ham}) is used with three different parameter sets:
\mbox{$\Delta\epsilon=0$} (black circles),
\mbox{$\Delta\epsilon=a_1$} (red triangles pointing down)
and \mbox{$\Delta\epsilon=-a_1$} (blue triangles pointing up).
\mbox{$B({\rm E2};J^\pi_{\rm i}\rightarrow J^\pi_{\rm f})$} is defined as
\mbox{$B({\rm E2};j^n2_1^+\rightarrow j^n0_1^+)$} for even $n$
and \mbox{$B({\rm E2};j^n\sfrac{11}{2}^+_1\rightarrow j^n\sfrac{7}{2}^+_1)$} for odd $n$,
in units $e^2b^4$,
where $e$ is the effective charge of the nucleon
and $b$ is the oscillator length.}
\label{f_d5g7} 
\end{figure}
In order to identify the important features that determine the odd--even E2 relation,
consider the schematic Hamiltonian
\begin{equation}
\hat H=\epsilon_{5/2}\hat n_{5/2}+\epsilon_{7/2}\hat n_{7/2}
-4\pi\sum_{T=0,1}a_T\sum_{i<j=1}^A\delta(\bar r_i-\bar r_j)\delta(r_i-R_0)C(R_0),
\label{e_ham}
\end{equation}
which includes the single-particle energies $\epsilon_{5/2}$ and $\epsilon_{7/2}$,
and a SDI with isoscalar and isovector strengths $a_T$,
and where $C(R_0)$ is a radial integral~\cite{Brussaard77}.
For nuclei with either neutrons or protons in the valence shell
the problem is independent of the isoscalar interaction
and eigenfunctions depend on a single parameter \mbox{$\Delta\epsilon/a_1$},
with \mbox{$\Delta\epsilon\equiv\epsilon_{5/2}-\epsilon_{7/2}$}.

The results of a schematic calculation are shown in Fig.~\ref{f_d5g7},
where \mbox{$B({\rm E2};j^n2_1^+\rightarrow j^n0_1^+)$} values in even--even nuclei
and \mbox{$B({\rm E2};j^n\sfrac{11}{2}^+_1\rightarrow j^n\sfrac{7}{2}^+_1)$} values in odd-mass nuclei
are plotted as a function of the nucleon number $n$
for three choices of the ratio \mbox{$\Delta\epsilon/a_1$}.
For degenerate orbitals, \mbox{$\Delta\epsilon=0$},
the even--even and odd-mass $B$(E2) values are connected by two shifted parabolas,
in agreement with seniority relations for several degenerate orbitals.
These relations are no longer valid if \mbox{$\Delta\epsilon\neq0$}
but it is seen that the E2 strength in the odd-mass nuclei
is much more affected than that in the even--even nuclei.
This introduces an odd--even staggering in the $B$(E2) values
that is more pronounced than in the case of degenerate orbitals.
The figure also illustrates the effect of particle--hole conjugation,
which for the Hamiltonian~(\ref{e_ham}) corresponds
to the replacement \mbox{$\Delta\epsilon\rightarrow-\Delta\epsilon$}.
For \mbox{$\Delta\epsilon=0$} the Hamiltonian~(\ref{e_ham})
is invariant under particle--hole conjugation
and the E2 transitions for the systems with $n$ and \mbox{$\Omega-n$} nucleons are identical.
Furthermore, E2 transitions in the $n$-nucleon system with \mbox{$\Delta\epsilon$}
are identical to those in the \mbox{$(\Omega-n)$}-nucleon system with \mbox{$-\Delta\epsilon$}.

\begin{figure}
\centering
\includegraphics[width=7cm]{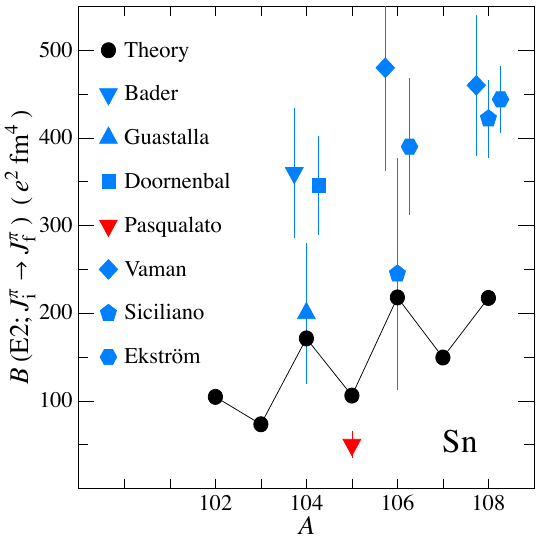}
\caption{Measured $B$(E2) values (coloured polygons) in the neutron-deficient tin isotopes
compared to theoretical results (black dots) obtained with the schematic Hamiltonian~(\ref{e_ham})
with the parameters \mbox{$a_1=0.35$}~MeV and \mbox{$\epsilon_{5/2}-\epsilon_{7/2}=-0.5$}~MeV.
\mbox{$B({\rm E2};J^\pi_{\rm i}\rightarrow J^\pi_{\rm f})$} is defined as
\mbox{$B({\rm E2};j^n2_1^+\rightarrow j^n0_1^+)$} for even $n$
and \mbox{$B({\rm E2};j^n\sfrac{11}{2}^+_1\rightarrow j^n\sfrac{7}{2}^+_1)$} for odd $n$,
in units $e^2{\rm fm}^4$.
Data are taken from
Refs.~\cite{Vaman07,Ekstrom08,Guastalla13,Bader13,Doornenbal14,Siciliano20,Pasqualato23}.}
\label{f_e2sn} 
\end{figure}
Figure~\ref{f_e2sn} compares the relevant E2 data in the neutron-deficient tin isotopes
with the results calculated with the schematic Hamiltonian~(\ref{e_ham}).
As the shell-model space is very restricted,
reasonable results can only be expected for \mbox{$102\leq A\leq108$}
since for larger mass numbers the orbitals $2s_{1/2}$, $1d_{3/2}$ and $0h_{11/2}$
in the upper half of the \mbox{50--82} shell cannot be neglected.
The parameters of the Hamiltonian are estimated from the energy spectra,
\mbox{$a_1=0.35$}~MeV and \mbox{$\epsilon_{7/2}-\epsilon_{5/2}=0.5$}~MeV.
The former parameter is adjusted to the spectra of $^{104}$Sn and $^{106}$Sn
and the latter reproduces approximately the $(\sfrac{7}{2}^+)$--$(\sfrac{5}{2}^+)$ splitting in $^{105}$Sn.
It is seen that the \mbox{$B({\rm E2};\sfrac{11}{2}^+_1\rightarrow\sfrac{7}{2}^+_1)$} value
measured in $^{105}$Sn is much smaller than the \mbox{$B({\rm E2};2_1^+\rightarrow0_1^+)$} values
in the surrounding even--even isotopes.
This finding is in line with the preceding theoretical analysis
although it seems that, the wide scatter of $B$(E2) values in the even--even isotopes notwithstanding,
the suppression of observed E2 strength in $^{105}$Sn is stronger than theoretically predicted.

\section{Conclusion}
\label{s_con}
The particle-core coupling model of an odd-mass nucleus,
which assumes a nucleon weakly coupled to an even--even core,
predicts \mbox{$B({\rm E2};J_{\rm i}\rightarrow J_{\rm f})$} values to the ground state of the odd-mass nucleus
that are equal to the \mbox{$B({\rm E2};2_1^+\rightarrow0_1^+)$} value in the core,
see Eq.~(\ref{e_relbe2}).
Under the assumption of identical nucleons ({\it i.e.}, either neutrons or protons) in a single-$j$ orbital
interacting through a seniority-conserving force,
a simple relation can be derived
that links the odd-mass \mbox{$J_{\rm i}\rightarrow J_{\rm f}$} E2 transitions
to the \mbox{$B({\rm E2};2_1^+\rightarrow0_1^+)$} value of the neighbouring even--even nucleus.
This shows that Eq.~(\ref{e_relbe2}) is valid only in the large-$j$ limit 
and that important deviations occur for finite values of $j$.
The deviations can be understood as a consequence of the Pauli principle,
which imposes anti-symmetry between the odd particle and the nucleons in the even--even core.
If the odd particle's type is different from that of the other valence nucleons
(case not discussed in this contribution),
deviations from the relation~(\ref{e_relbe2}) have a different origin
that can be traced to an increased collectivity in the odd-mass nucleus
due to the neutron-proton interaction.

The results presented here
can be extended to identical nucleons in several orbitals.
If the orbitals are degenerate in energy,
one recovers the single-$j$ result:
as a function of particle number $n$
the even--even and odd-mass $B$(E2) values
are connected by two parabolas, shifted with respect to each other.
The non-degeneracy of single-particle energies, however, has a major impact on this finding.
In the example that is relevant to the neutron-deficient tin isotopes,
namely neutrons in the \mbox{$1d_{5/2}+0g_{7/2}$} orbitals,
it is found that the odd-mass \mbox{$B({\rm E2};j^n\sfrac{11}{2}^+_1\rightarrow j^n\sfrac{7}{2}^+_1)$} value
can be strongly suppressed 
while the even--even \mbox{$B({\rm E2};2_1^+\rightarrow0_1^+)$} value
is less affected by the non-degeneracy of the single-particle energies.
This finding agrees qualitatively with the E2 strength
observed in neutron-deficient tin isotopes.

Measurements of E2 transition probabilities with smaller uncertainties
in both even--even and odd-mass neutron-deficient tin nuclei
would be very much welcome to further our understanding of the odd--even E2 relation.

\end{document}